# Crystal excitations features in the photon emission spectrum of the quantum channeled particle

E.A. Mazur

Department of Physics, National Research Nuclear University «MEPHI», Moscow, Kashirskoe shosse, 31, Russia

The objective of present paper is to examine the role of several previously not discussed diagrams that allowes us to describe the generation of the longitudinal and transverse collective excitations, in particular, phonons and plasmons, in the crystal with the channeled particle (CP) as well as the processes of the channeled-particle radiation with simultaneous generation of the scalar collective excitations in the crystal. We will not touch in the present consideration the main diagrams corresponding channeled-particle radiation. This processes of the channeled-particle radiation were thoroughly studied in the works of many authors (see $[1-8]$ and references therein) and are not subject to any revision. To address this issue we have to enrich the well-developed theory of radiation of the channeled particle with the account of the interaction of the channeling particle with the collective scalar excitations in the crystal along with the generation of photons. Let us consider the channeling particle with quasibloch behavior in the oriented crystal. The hamiltonian of the system is thought of as having three main terms $\hat{H} = \hat{H}_{cr} + \hat{H}_{op} + \hat{H}_{int}$ where $\hat{H}_{cr}$ is the crystal hamiltonian creating the potential $U(\vec{r})$ with $\hat{H}_{int} = eU(\vec{r})$ so that the equation for the unperturbed wave function $\psi_{n,\vec{p}}(\vec{r})$ of the fast oriented or channeled particle in the state with the quantum numbers $n, \vec{p}$ with neglect of the small gradient and quadratic in potential $U$ terms has the well known form (see, for example, $[1-8]$). We will consider the particle channeled within the distinct channels, $\theta$ is the angle of incidence of the relativistic particle with respect to the crystallographic planes, $d/\theta_L \sim 1$ so that coherent length $l \sim d/\theta_L$, $E \gg mc^2$, $U \sim 20 - 100\ eV$, d is the interplane distance. Let us consider the S-matrix of interaction $V = eU = Ze\int jA d^4x$ of 4-current $j = \{j_0, \vec{j}\}$ of the fast oriented charged particle with the 4-potential of excitations $A = \{\varphi, \vec{A}\}$ in the crystal. The probability of the process at the $t = +\infty$ should be defined as $W_{if} = \overline{|S_{if}|^2} \delta(E_i - E_f - \hbar\omega)$. Let us primarily consider processes to the first order in perturbing potential $j(x)A(x)$. For the first order probability after thermodynamic averaging over the crystal states we obtain the following formula

$$dW_{if}^{(1)} = e^2 \int j_{if}^{*(0)}(x) j_{if}^{(0)}(x') \langle T\Phi(x)\Phi(x')\rangle d^4x d^4x' dv + + e^2 \int j_{if}^{*(\alpha)}(x) j_{if}^{(\beta)}(x') \times$$
$$\langle TA_\alpha(x)A_\beta(x')\rangle d^4(x)d^4x' dv + e^2 \int j_{if}^{*(0)}(x) j_{if}^{(\alpha)}(x') \langle T[\Phi(x)A_\alpha(x')]_+\rangle d^4x d^4x' dv. \quad (1)$$

In (1) $\vec{j}_{if} = \overline{\psi_i(x)}\vec{\alpha}\psi_f(\vec{x})$ is the matrix element of the fast particle 4-current, $\vec{\alpha} = \gamma^0\vec{\gamma}$, $\gamma^\mu (\mu = 1,2,3)$, $\vec{A}$ is the vector component of radiation field so that with $D_{\mu\nu}(x,x') = i\langle TA_\mu(x)A_\nu(x')\rangle$ should be defined the photon correlators in the crystal. We will not consider the second term in (3) because this term describes the processes of the channeling particle radiation the theory of which is well developed [1–8] to date. Let us at first investigate the role of the first scalar term in (2) describing the phonon and plasmon generation with the fast oriented or channeled particle in the crystal. In this formula the resulting perturbing potential of the electron and ion subsystems is written down as $\Phi(\vec{q}+\vec{G},t) = V_c(\vec{q}+\vec{G})\delta\rho_e(\vec{q}+\vec{G},t) + \sum_p V_p(\vec{q}+\vec{G})\delta\rho_p(\vec{q}+\vec{G},t)$. Here $\vec{G}$ is the reciprocal lattice vector of the crystal. When substituting $\Phi$ into (2) the first scalar term in (2) responding to the probability of Cherencov generation of collective electron $W_{if}(electron)$ and phonon $W_{if}(phonon)$ excitations with the energy $\omega$ by the channeling particle is shown to have the form $W_{if} = W_{if}(electron) + W_{if}(phonon) + W_{if}(int)$ with the intereference term $W_{if}(int)$ in it. The probability of the process of collective electron excitations with the oriented or channeled particle $W = \sum_q W_{if} E_i / pv$ is obtained with $W_{if}$ in it having the following form

$$dW_{if} = \sum_{\vec{G}} \frac{e^2 \, \mathrm{Im}\, \varepsilon^{-1}(\vec{q},\vec{q}+\vec{G},\omega)}{q^2[1-\exp(-\hbar\omega/T)]} \langle f|e^{i\vec{q}\vec{r}}|i\rangle \langle i|e^{-i(\vec{q}+\vec{G})\vec{r}}|f\rangle \delta(E_i - E_f - \gamma\hbar\omega). \quad (2)$$

In (3) $\varepsilon^{-1}(\vec{q},\vec{q}+\vec{G},\omega)$ is the inverse matrix dielectric function of the electrons in the crystal, p and v are the impulse and velocity of the fast particle respectively, $\gamma$ is the Lorentz-factor for the particle. Plasmon dispersion $\omega(\vec{q})$ is determined with the pole of the inverse dielectric function $\varepsilon(\vec{q},\omega)$ so that $\mathrm{Im}\left[-\frac{1}{\varepsilon(\vec{q},\omega)}\right] = \frac{\pi\omega_p^2}{2\omega^2(\vec{q})}\delta(\omega-\omega(\vec{q}))$ where $\omega^2(\vec{q}) = \omega_p^2 + \left(\frac{q^2}{2m}\right)^2$. The probability of the phonon generation by the channeled particle has the following form

$$W_{if}(phonon) = e^2 \sum_{\vec{G},\vec{G}'} \int\int d^3\vec{x}\, d^3\vec{x}' \left(\psi_f^* \exp(i(\vec{q}+\vec{G})x)\right)\psi_i(\vec{x})\left(\psi_i^*(\vec{x}')\exp(i(\vec{q}+\vec{G}')x')\right) \times$$
$$\times \psi_f(\vec{x}') \sum_{pp'} V_p(\vec{q}+\vec{G}) V_p(\vec{q}+\vec{G}') S_{pp'}(\vec{q}+\vec{G},\vec{q}+\vec{G}',\omega)\delta(E_i - E_f - \gamma\hbar\omega). \quad (3)$$

In (3) $V_p(\vec{q}+\vec{G},\omega) = \sum_{\vec{G}'} V_{cp}(\vec{q}+\vec{G})\varepsilon^{-1}(\vec{q}+\vec{G},\vec{q}+\vec{G}',\omega)$ is the dynamically screened type-p ion potential in the crystal, $S_{kp}(\vec{q}+\vec{G},\vec{q}+\vec{G}',\omega) = S_{kp}^{(1)} + S_{kp}^{(2)} + S^{(int)} + \ldots$. One phonon dynamical structure factor $S_{kp}^{(1)}(\vec{q}+\vec{G},\vec{q}+\vec{G}',\omega)$ is represented with the use of the partial k-type Debye-Waller factor of the crystal lattice $Y_k(\vec{q}+\vec{G}'-\vec{G}'')$ in the following form

$$S_{kp}^{(1)}(\vec{q}+\vec{G},\vec{q}+\vec{G}',\omega) = \exp\left[-Y_k(\vec{q}+\vec{G}'-\vec{G}') - Y_p(\vec{q}+\vec{G}'-\vec{G}')\right]\sum_\lambda F_k^{(1)}(\lambda) F_p^{*(1)}(\lambda) \times$$
$$\times \exp\left[-i(\vec{q}+\vec{G}'-\vec{G}'')\vec{R}_{kp}^{(0)}\right]\sum_{\pm} \Delta(\vec{q}_\lambda \pm q)\left[n(\lambda) + \frac{1}{2} \mp \frac{1}{2}\right]\delta(\omega \pm \omega(\lambda)). \quad (4)$$

Here $\vec{R}_{kp}^{(0)} = \vec{R}^{(0)}(0,k) - \vec{R}^{(0)}(0,p)$, $\vec{R}^{(0)}(0,k)$ - radius vector of the k-type ion with the mass $M_k$ in the elementary crystal cell, $\vec{e}(k,\lambda)$ is the wave vector $\lambda$ dependent polarisation vector of the k-type phonon, $F_k^{(1)}(\lambda) = (\hbar/2M_k N\omega(\lambda))^{1/2}\left[(\vec{q}+\vec{G}')\vec{e}(k,\lambda)\right]$, N is the number of cells in the crystal, $n(\lambda)$ is the Bose distribution function. Two-phonon and multiphonon dynamical structure factors should be written down in an analogous manner. And now let us proceed to the processes of the second order. Let us consider the mutual influence of the processes of radiation and generation of excitations inside the quantum crystal with the channeled particle (electron or positron). Turning to the second order processes and substituting in $S^{(2)}$ the wave functions $\psi(x) = \psi(\vec{r})\exp\left[-i\left(E - \frac{i}{2}\Gamma\right)t\right]$ of the channeled particle in the form that takes into account the decay of the quantum states of the CP with time t we can derive the part $S^{(2)}$ of the S-matrix with the use of the Green fuction $S_c(x',x)$ of the CP in the intermediate state [10] in the form wherein the first part of $S^{(2)}$ describes the process of the CP interaction with the scalar potential $A_0(x)$ of an electron excitation in the crystal with the subsequent interaction with the vector potential $A_\beta(x')$ of the gamma quantum (first type process). The second part of $S^{(2)}$ describes the process of the CP interaction with the inverse in time order of interaction of the CP with the electron excitation in the crystal and the gamma quantum (second type process). For the first type process we obtain for the CP transfer probability $dW_{if}^{(2)'}$ after summation over the photon polarizations the following expression for the planar channeling

$$dW_{if}^{(2)'} = \pi e^2 \sum_{\omega,\omega'}\sum_{q_x,q_{x'}} |V_{is}(q_x)|^2 \left|\langle s|\exp[iq_x'x']|f\rangle\right|^2 \frac{\gamma^2}{(E_s - E_i + \gamma\omega)^2 + (\Gamma_s + \Gamma_i)^2/4} \quad (5)$$
$$\times S(\vec{q},\omega)D(\vec{q}',\omega')\delta(q_{fy} + q_y' - q_{iy} + q_y)\delta(q_{fz} + q_z' - q_{iz} + q_z)\delta(E_f - E_i - \gamma\omega - \omega').$$

In (5) $D(\vec{q},\omega')$ is the photon correlator. An analogous expression should be written for the second type process $dW_{if}^{(2)''}$ probability. The cross section of the process should be calculated with averaging of $dW_{if}^{(2)'}$ on the polarisations of the channeled particle and with integration of $dW_{if}^{(2)'}$ on all the gamma quantum impulses and electron excitation impulses in the crystal relevant given energy of the photon and electron excitation. In the resonance conditions when $E_s - E + \gamma\omega = 0$ the probability of such compound process is scaled with the probability of the single photon resonant emission with the evident from (5) parameter $4\gamma^2 V^2/(\Gamma_s + \Gamma)^2$ which can easily be matched to or greater than unity. The same situation is with the generation by the channeled particle of all excitations leading to the poles in the electron or phonon dynamical structure factor of the crystal. In other words, all the pole – like features of the dynamical crystal structure are displayed in the form of additional series of channeled particle radiation [8,9]. Substituting in (5) the expression for the one-phonon dynamical structure factor $S_{kp}^{(1)}(\vec{q}+\vec{G},\vec{q}+\vec{G}',\omega)$ with the energy $\omega$ of the excitation in the lab or for the two-phonon dynamical srtucture factor $S_{kp}^{(2)}(\vec{q}+\vec{G}',\vec{q}+\vec{G}'',\omega)$ and calculating the integral over pole singularity in $\omega$ we get the cross sections of the process of the longitudinal phonon emission by the channeled particle accompanied with the gamma quantum radiation. Substituting in (5) the dynamical structure factor that describes the collective electron excitation in the crystal $S(\vec{q},\omega)=\pi\delta(\omega-\omega(\vec{q}))\omega_p^2/2\omega^2(\vec{q})$ we obtain after integration in $\omega$ over the pole the cross section of the process of the collective electron excitation emission by the channeled particle accompanied with the gamma quantum radiation. Inserting in (5) the known expression for the photon Green function $D_{c\alpha\beta}(x-x')$ [10] and integrating over the pole in $\omega'$ we easily obtain the probability of the combined process under consideration

$$dW_{if}^{(2)'} = \frac{\pi^2}{2}e^2 \sum_{\omega',q_x'} \left\langle \left|V_{is}(q_x)\right|^2 \right\rangle_{AV} \left|\left\langle s \left|\exp\left[iq_x'x'\right]\right| f \right\rangle\right|^2 \frac{\gamma^2/\omega'}{(E_s-E_i+\gamma\omega_{pl})^2+(\Gamma_s+\Gamma_i)^2/4} \quad (6)$$

$$\times \delta(q_{fy}+q_y'-q_{iy})\delta(q_{fz}+q_z'-q_{iz})\delta(E_i-E_f \mp \gamma\omega_{pl}-\omega').$$

Energy-momentum conservation laws for the quantum channeled particle are obtained with the use of the Bloch conditions for the CP. So within $\gamma \gg 1$ the result for the photon energy with the use of the impulse relations has the following form which also takes into account the refractive index of the crystal [2] $n(\omega)$ $\hbar\omega = \left(\Delta E_\perp' \sqrt{1-(v_\parallel/c)^2} \mp \hbar\omega_{pe}\right)/(1-(v_\parallel/c)n(\omega)\cos\theta)$. Essentially, this formula represents the record of the Doppler effect for the process of simultaneous emission of a photon and a plasmon with the channeled particle. The resulting emission peaks have a large half-width due to the plasmon momentum carryover, in contrast to the channeled particle radiation process [1−8] without plasmon emission. Thus a cone of the emitted photons undergoes blurring. The photon energy in the process of the phonon emission combined with photon emission can be written down as

$\hbar\omega = \left(\Delta E_\perp' \sqrt{1-(v_\parallel/c)^2} \pm m\hbar\omega_{phonon}\right)/(1-(v_\parallel/c)n(\omega)\cos\theta)$. The radiative processes without altering the quantum level of transverse motion (that may be called channeling particle quantum fluorescence) are illustrated with the processes of plasmon or phonon conversion. Radiation of the photon after the plasmon absorption is characterized with the energy of the emitted gamma quantum

$\hbar\omega = \dfrac{\hbar\omega_{pe}}{1-(v_\parallel/c)n(\omega)\cos\theta}$. The process of radiation of the photon after the phonon absorbtion is characterized with the following energy of the gamma-quantum $\hbar\omega = \dfrac{m\hbar\omega_{transv1,phonon} \pm s\hbar\omega_{transv2,phonon}}{1-(v_\parallel/c)n(\omega)\cos\theta}$.

The radiative processes with altering the quantum level of the transverse motion with photon energy

$\hbar\omega = \dfrac{\Delta E_\perp' \sqrt{1-(v_\parallel/c)^2} \pm m\hbar\omega_{phonon}}{1-(v_\parallel/c)n(\omega)\cos\theta}$ and $\hbar\omega = \dfrac{\Delta E_\perp' \sqrt{1-(v_\parallel/c)^2} \pm \hbar\omega_{pe}}{1-(v_\parallel/c)n(\omega)\cos\theta}$ respectively can be named as

multiphonon and plasmon "wings" in radiation. Conditions for the experimental observation of the channeling plasmon-photon or multiphonon-photon radiation peaks are summarized as follows: 1. Potential in lab. system: $U_0 \sim 20 eV$; $d \sim 0.2$-$0.3$ Å (Si); in related system ($v_x = 0$): $U = U_0 (E/mc^2)$. 2. Number of levels within the potential wall: $N \sim P_{z\,max}d/h \sim (EU_0)^{1/2}d/hc$. 3. Distance between levels in related system: $\Delta E \sim U/N \sim (EU_0)^{1/2}(h/mcd)$. 4. In resonance conditions quantum fluorescence effect and radiation "wings" will be very noticeable on the phone of the bremsstrahlung ; 5. Resonance can be reached with adjusting the energy of the channeling particle and correctly orienting CP beam if

$E/U_0 > (\lambda_0/2d)^2 \sim 10^{7-8}$. The results obtained by experimenters have not yet discovered, due to both a decrease in the intensity of the peaks for the «wings» of radiation and for the conversion and a higher frequency range, where they are to be found. In addition to the processes studied in the present work known [4,8] coherent elastic processes of the interaction of the fast particle with the periodic deviation of the crystal potential from its averaged values are emerging with the appearance of periodic in time perturbations $Ve^{i\omega_t t}$ with the period $T = \pi\hbar/V_{mn}$ in traveling of the fast particle through the crystal.

**Conclusions**: Expressions are obtained for the probability of emission of phonons and plasmons with the channeled particle. The description of the processes of the crystal excitation with the channeled particle accompanied by the simultaneous emission of a photon is developed. It is proved the previously predicted effect [8,9] that all the specific features of the electron and phonon crystal excitation structure appear as components of radiation of the oriented fast charged particle. The emergence of new peaks in the emission spectrum of such channeled particle associated with the processes of simultaneous resonant hard gamma-quantum and plasmon (phonons) excitation during its motion in the crystal is discussed. The distance between the particle transverse motion levels should be equal to the sum of the photon energy and quantum plasmon energy in the coordinate system associated with the moving particle. It is found that the photon-plasmon radiation effect probability is of the same order of magnitude with a known standard radiation process probability.

References

[1] N.P.Kalashnikov, Coherent Interactions of the Charged Particles in Momocrystals, Atomizdat, Moscow, 1981 (in Russian).

[2] V.G.Baryshevsky, Channeling, radiation and reactions in cristals at high energy (in Russian). M., Moscow Iniversity Press, 1982.

[3] V.A. Bazylev, N.K. Zhevago, Radiation of Fast Particles in Matter and External Fields, Nauka, Moscow, 1987 (in Russian).

[4] V.V.Beloshitskii, S.B.Dabagov. Zh.Tekh.Fiz. 58 (1988) 1563.

[5] M.A. Kumakhov and R. Weddel, Radiation of relativistic light Particles during Interaction with Single Crystals. Heidelberg, Spectrum, 1991.

[6] V.N.Baier, V.M.Katkov, V.M.Strakhovenko, Electromagnetic Processes at High Energies in Oriented Single Crystals, Novosibirsk, Nauka, 1989 (in Russian); Singapore World Scientific, 1998 (in English).

[7] A.I.Ahiezer, N.F. Shulga, Electrodynamics of high-energy particles in matter (in Russian), M., Nauka, 1993.

[8] S.B.Dabagov, N.K.Zhevago. Riv. del Nuovo Cim. 31(2008) 491.

[9] E.A.Mazur. Theses of reports of 17-th All-Union Conference on the Physics of interaction of charged particles with crystals. (1987) 99, Manifestation of the crystal excitations energy spectrum features in radiation of oriented particle.

[10] A.I.Ahiezer, V.B.Berestecky, Quantum Electrodynamics, Nauka, Moscow, 1981 (in Russian).